\documentclass[conference,10 pt]{IEEEtran}
\usepackage{graphicx}
\usepackage{xspace}
\usepackage{hyperref}
\usepackage{amssymb}
\usepackage{wrapfig} 
\usepackage{booktabs}
\usepackage{pgfplots}
\usetikzlibrary{backgrounds}
\newcommand\eclause{e-clause\xspace}
\newcommand\eclauses{e-clauses\xspace}
\newcommand\tool[1]{{\sc #1\xspace}}
\renewcommand\c[1]{\mbox{(#1)}}

\newcommand\Long[1]{#1}
\newcommand\Short[1]{}

\usepackage{amsthm}

\newtheoremstyle{ieeeconf}
{0pt}   
{0pt}   
{\normalfont}  
{\parindent}       
{\itshape} 
{:}         
{ } 
{\thmname{#1} \thmnumber{#2}\thmnote{ (#3)}} 
\makeatletter
\renewenvironment{proof}[1][\proofname]{\par
	\pushQED{\qed}%
	\normalfont \topsep\z@
	\trivlist
	\item[\hskip2em
	\itshape
	#1\@addpunct{:}]\ignorespaces
}{%
	\popQED\endtrivlist\@endpefalse
}
\makeatletter

\theoremstyle{ieeeconf}

\newtheorem{definition}{Definition}
\newtheorem{example}{Example}
\newtheorem{proposition}{Proposition}
\newtheorem{observation}{Observation} 
\newcommand\email[1]{{\tt #1}}
\begin{document}
\title{Exploiting Isomorphic Subgraphs in SAT \Long{\\ (Long version)}}

\author{
	\IEEEauthorblockN{Alexander Ivrii} \IEEEauthorblockA{IBM Haifa Research Lab, Israel \\
		\email{alexi@il.ibm.com} } \and
	\IEEEauthorblockN{Ofer Strichman} \IEEEauthorblockA{Information
		System Engineering, \\ IE, Technion, Haifa, Israel \\
		\email{ofers@ie.technion.ac.il} }
} 	
\maketitle

\begin{abstract}
  While static symmetry breaking has been explored in the SAT community for decades, only as of 2010 research has focused on exploiting the same discovered symmetry dynamically, during the run of the SAT solver, by learning extra clauses. The two methods are distinct and not compatible. The former may prune solutions, whereas the latter does not -- it only prunes areas of the search that are guaranteed not to have solutions, like standard conflict clauses. Both approaches, however, require what we call \emph{full symmetry}, namely a propositionally-consistent mapping $\sigma$ between the literals, such that $\sigma(\varphi) \equiv \varphi$, where here $\equiv$ means syntactic equivalence modulo clause ordering and literal ordering within the clauses. In this article we show that such full symmetry is not a necessary condition for adding extra clauses: isomorphism between possibly-overlapping subgraphs of the colored incidence graph is sufficient. While finding such subgraphs is a computationally hard problem, there are many cases in which they can be detected a priori by analyzing the high-level structure of the problem from which the CNF was derived. We demonstrate this principle with several well-known problems\Long{, including Van der Waerden numbers, bounded model checking and Boolean Pythagorean triples}.
\end{abstract}

\section{Introduction: symmetry, almost symmetry, and \eclauses}\label{sec:intro}
Symmetry breaking~\cite{S09} is a well known technique for accelerating SAT solving, which originated decades ago by Puget~\cite{P93} for CSP, and later by Crawford et al.~\cite{CGLR96} for CNF. Symmetry-breaking for CNF was implemented efficiently in the tool \tool{shatter}~\cite{ARMS03} and later improved in \tool{BreakID} \cite{DBBD16}. In a nutshell, it means that new predicates, called \emph{symmetry-breaking} predicates, are added to the input formula $\varphi$, without changing its satisfiability. These predicates prune the search space and are likely to remove solutions, but without changing the satisfiability of the formula. The construction of those predicates is based on finding a mapping $\sigma$ between the literals of the input formula $\varphi$, such that $\sigma(\varphi) \equiv \varphi$. Here `$\equiv$' means syntactic equivalence modulo clause ordering and literal ordering within the clauses. The mapping has to be \emph{propositionally-consistent}, which means that $\forall v_1,v_2 \in var(\varphi).\ \sigma(v_1) = v_2 \Rightarrow \sigma(\bar{v}_1) = \bar{v}_2$ and $\sigma(v_1) = \bar{v}_2 \Rightarrow \sigma(\bar{v}_1) = v_2$.
If we find such a mapping, then it means that every satisfying solution $\alpha$ to $\varphi$ has the property that $\sigma(\alpha)$ also satisfies $\varphi$. We can then add a constraint that prunes one of those solutions. As an example, consider
\[\varphi = \c{1 -3} \c{2 -3} \c{1 2 3} \c{-1 -2}\]
and the mapping $\sigma: {1\mapsto 2, 2 \mapsto 1}$ (by convention, each such mapping implies that the mapping of the negated literals is also included in $\sigma$, e.g., $-1\mapsto -2 \in \sigma$). We see that
\[ \sigma(\varphi) = \c{2 -3}\c{1 -3}\c{2 1 3}\c{-2 -1}\;,\]
and that $\sigma(\varphi)\equiv \varphi$. Indeed if we take any solution $\alpha$ to $\varphi$, we see that $\sigma(\alpha)$ is a solution as well. For example, for $\alpha = (1,2,3) \mapsto(T,F,F)$ we have $\alpha\models \varphi$, and indeed $\sigma(\alpha) \models \varphi$ as well, since $\sigma(\alpha) = (1,2,3)\mapsto (F,T,F)$. Crawford et al. showed how to add symmetry-breaking constraints, which we will not detail here. In this case it may amount to adding the clause $\c{-1 2}$, which indeed in this case excludes the first solution without excluding the second one. \Short{Such pruning of solutions is in many cases helpful for shortening the overall run-time~\cite{ARMS03,K15}.}

Symmetry-breaking tools discover such mappings by analyzing the colored literals incidence graph\footnote{Such a graph is constructed from a CNF by introducing a vertex for each literal and each clause, connecting opposite literals with an edge, and connecting the literals to the clauses that they are part of. The clauses' nodes have one color, and the literals' nodes have a different color.} $G$ with respect to multiple potential mappings $\Sigma$:  if for $\sigma \in \Sigma$ it holds that $\sigma(G)\equiv G$ (this is called `automorphism'), then $\sigma$ defines a symmetry. The isomorphism in this case is restricted such that for every two nodes, $n_1,n_2 \in G$, if $\sigma(n_1) = n_2$ then $n_1$ and $n_2$ must have the same color, i.e., clause nodes are mapped to clause nodes and literal nodes to literal nodes.

\Long{Symmetry breaking can be effective in shortening the run time if it helps the solver skip the inference process that is required for detecting that certain partial solutions cannot be extended to full solutions. This is especially relevant when there are many such un-extendable partial assignment that are pruned by those predicates, whether the formula is satisfiable or not. Symmetry breaking can also slow down the solver, for example when the formula is satisfiable and it happens that the solution that is still possible after adding those predicates is harder to find than those that were excluded.
Indeed, symmetry breaking has been shown to be effective especially in unsatisfiable formulas, and enabled, for example, an increase in the size of `pigeonhole' problems (which are unsat) that can be solved within a given time limit~\cite{ARMS03,K15}.}

Another way to exploit symmetry is by adding clauses during search. Henceforth we will call such clauses `\eclauses', for `Extra' clauses.
This option has mostly been researched in the CSP community, under the names \emph{Symmetry breaking during search - SBDS} ~\cite{BW02,GHK02,GS00,CBMS14} and \emph{Symmetry Breaking by Dominance Detection - SBDD}~\cite{FSS01}. In the SAT community this route was first explored via the Symmetrical Learning Scheme (SLS)~\cite{BNOS10}, which adds new clauses during the search based on learned clauses and a pre-computed set of symmetry `generators'. SLS was later improved by Symmetry Propagation (SP)~\cite{DBCDM12}, which only adds such extra clauses if they lead to further (immediate) propagations, and
several years later by Symmetric Explanation Learning (SEL)~\cite{DBB17}, which is integrated within BCP (it takes the reason clause of the propagation as the base for adding \eclauses). According to~\cite{DBB17}, SEL is the only one of those that is competitive with modern static symmetry breaking. Finally,~\cite{TD19} has a similar scheme in which \eclauses are only added if the learned clause has a low LBD. In~\cite{DBB17} those methods were jointly called \emph{dynamic} symmetry \emph{handling}, to emphasize that unlike \emph{static} symmetry \emph{breaking} they are based on an analysis during the search (hence `dynamic'), and that they do \emph{not break symmetry}, as they do not remove solutions.
We find this name inadequate, however, because symmetry does not need to be `handled'. A more proper name is dynamic symmetry \emph{exploitation}, which is the name we will use in the rest of this article. Although static symmetry breaking and dynamic symmetry exploitation are based on the same data -- the symmetries in the formula -- they are not compatible. One cannot use dynamic symmetry exploitation if the symmetries it relies on are broken by added predicates.

Dynamic symmetry exploitation was also studied for the case of \emph{almost symmetric} formulas (also called `weak symmetry')~\cite{M05,CBMS14}, formalized as follows. Let
\begin{equation} \label{eq:almost}
	\varphi \equiv\varphi_1\cup\varphi_2\;,
\end{equation}
where here we equate formulas $\varphi, \varphi_1,\varphi_2$ with sets of clauses. Let $\sigma$ be a literal map of $\varphi$ such that
\begin{equation}\label{eq:e22}
	\sigma(\varphi_2 )\equiv\varphi_2\;.
\end{equation}
This reflects a common scenario, where a few clauses -- marked here by $\varphi_1$ -- disrupt the symmetries in the formula.  
The main method that was suggested in these references is to add \eclauses based on $\varphi_2$. That is, once a clause $c$ is learned from $\varphi_2$ alone, add $\sigma(c)$ as well.

In this article we observe that the requirement of symmetry as used by all of those prior works on dynamic symmetry exploitation is a sufficient, yet not a \emph{necessary} condition for adding \eclauses. 
We will need the following definitions for explaining this claim. 
\begin{definition}[The \emph{refined} colored incidence graph]
The \emph{refined} version of a colored incidence graph assigns separate colors to clauses of different arity. 
\end{definition}
We will denote this graph by $G$, assuming the underlying formula is clear from the context (it can also include learned clauses).
\Long{
\begin{definition}[The subgraph induced by a clause]\label{def:induced}
Given a clause $c$, the subgraph of $G$ that is \emph{induced} by $c$ is comprised of the nodes in $G$ that correspond to $c$ and its literals, and the edges between them. 
\end{definition}
We now overload this term so it can be used with a resolution sequence: 
\begin{definition}[The subgraph induced by a resolution sequence]\label{def:seq}
Given a resolution sequence $c_1,\ldots, c_n$, its corresponding \emph{induced} subgraph in $G$ is comprised of the subgraphs induced by its clauses (Def.~\ref{def:induced}) and the edges between opposite literals that were resolved in the sequence. 
\end{definition}
}
\Short{
\begin{definition}[The subgraph induced by a resolution sequence]\label{def:seq}
	Given a resolution sequence $c_1,\ldots, c_n$, its corresponding \emph{induced} subgraph in $G$ is comprised of the subgraphs induced by these clauses, and the edges between opposite literals that were resolved in the sequence. 
\end{definition}
}
Now, consider such a resolution sequence $c_1,\ldots, c_n$ that was used for learning a clause $c$ ($c$ itself is not part of the sequence), and its corresponding induced subgraph $g$. Consider also another subgraph $g'$ of $G$ that is color-isomorphic to $g$. It is not hard to see that $g'$ reflects another possible resolution sequence in the formula, ending with a different clause, which we can add as an \eclause. 
This criterion is \emph{ad-hoc} and does not require automorphism of the original formula or some pre-defined part of it as in almost-symmetries. In fact, it can be seen as an application of the SR-II inference rule suggested by Krishnamurthy in~\cite{K85} already in 1985 (there was no indication, however, how a solver may exploit that rule in~\cite{K85}).
In some types of formulas, finding \eclauses based on this reasoning is computationally cheap, and can lead to improvements in the overall run-time of the solver. 
The important point is that this technique can be applied even when there is no mapping $\sigma$ such that  $\sigma(\varphi)\equiv \varphi$, which implies that this technique can derive \eclauses that cannot be derived by the above-mentioned symmetry exploitation techniques.  

In fact, this idea was implicitly used in the past by the second author~\cite{S00} for adding \eclauses in the case of bounded-model checking problems, and by Say et al. for adding such clauses in the case of optimizing a planning process with neural networks~\cite{SDNS20}. Both references reported performance gains. In this article we give a general view that encompasses also these two references, and show that the potential for such clauses is present in many other types of formulas.

\Long{Let us demonstrate this idea with an example.}
\begin{example} \label{ex:wd}
Let $\varphi$ be comprised of the following clauses:
\begin{equation}\label{eq:ex}
	\begin{tabular}{ll@{\hspace{0.7 cm}}ll@{\hspace{0.7 cm}}ll}
		\c{1 2 3} & \c{-1 -2 -3} & 
		\c{2 3 4} & \c{-2 -3 -4} \\ 
		\c{3 4 5} & \c{-3 -4 -5} &
		\c{4 5 6} & \c{-4 -5 -6} \\ 
		\c{5 6 7} & \c{-5 -6 -7} &
		\c{1 3 5} & \c{-1 -3 -5} \\
		\c{2 4 6} & \c{-2 -4 -6} &
		\c{3 5 7} & \c{-3 -5 -7} \\
		\c{1 4 7} & \c{-1 -4 -7} 
	\end{tabular}
\end{equation}
It happens to be the Van der Waerden formula (3,3; 7). We will describe this type of formulas later, in section~\ref{sec:vdw}.

Symmetry breaking, as emitted by \tool{BreakID}, discovers the two mappings below (these are also called `generators'). To get to the full set of possible mappings one needs to also consider their compositions.
\begin{equation} \label{eq:br}
\begin{array}{l}
\sigma_1:\ \mbox{[ 1 7 ] [ 2 6 ] [ 3 5 ]} \\
\sigma_2:\ \mbox{[ 1 -1 ] [ 2 -2 ] [ 3 -3 ] ... [ 7 -7 ]}
\end{array}
\end{equation}
This representation is called `cycle form', and should be interpreted as follows: in each line, every literal appears at most once; it should be replaced with the literal that comes next in the brackets, and if it is the last one then with the first literal in the brackets. In this example $\sigma_1$ implies that simultaneously swapping literals 1 and 7, 2 and 6, 3 and 5 (and correspondingly, their negated versions, -1 and -7, etc.) results in the same formula. Readers familiar with Van der Waerden formulas may notice that this symmetry corresponds to a reversal of the indices, i.e., the first variable becomes last, the second one becomes second to last, etc, and that $\sigma_2$ corresponds to a swap of the colors. In such formulas, regardless of their length, these are the only two possible symmetries. 

Now suppose that we learn a new conflict clause $c = \c{1 2 -5 6}$, via the following resolution sequence:
\begin{equation}\label{eq:res}
\mbox{(1 2 3), (-3 -4 -5), (2 4 6)}\;.
\end{equation}
We can therefore add two \eclauses corresponding to the two generators:
\begin{equation}
\sigma_1\c{1 2 -5 6} = \c{7 6 -3 2} \qquad \sigma_2\c{1 2 -5 6} = \c{-1 -2 5 -6}\;.
\end{equation}
However, more \eclauses can be derived based on this conflict clause. We need to find a subgraph of $G$ that is color-isomorphic to the one representing the sequence (\ref{eq:res}). 
Going back to our example, it is indeed not hard to see that (2 3 4), (-4 -5 -6), (3 5 7), all of which are clauses in $\varphi$, give us just that -- see Fig.~\ref{fig:iso}. Applying the same resolution steps yields a new \eclause (2 3 -6 7), which cannot be deduced by any composition of $\sigma_1,\sigma_2$, simply because our inference is not based on the original CNF's symmetry, rather it is inferred dynamically from the resolution process.
\hfill $\qed$
\begin{figure}
  \centering
  \begin{tabular}{lr}
  \scalebox{0.6}{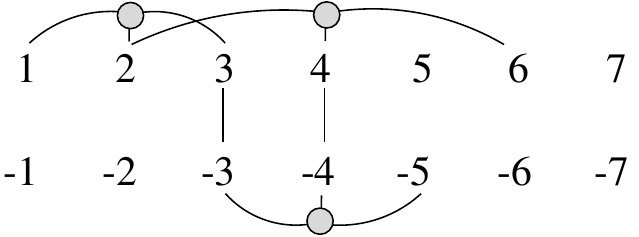}\hspace{0.8 cm} 
  \scalebox{0.6}{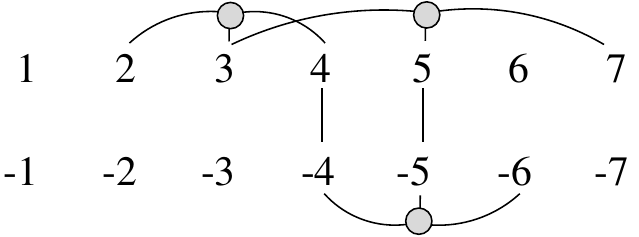}
  \end{tabular}
  \caption{Two isomorphic subgraphs of the same refined colored incidence graph corresponding to (\ref{eq:ex}). The literals are nodes with a separate color than the clause nodes. All the clause nodes in this example are of the same arity, hence they have the same color.  }\label{fig:iso}
\end{figure}
\end{example}

\Long{
\begin{example}
In the Boolean Pythagorean triples problem with 2 colors, we attempt to assign one of two colors to each number up to a given $n \in \mathbb{N}$, such that no Pythagorean triple (i.e., three numbers  $a,b,c$ that are related by $a^2 + b^2 = c^2$) is monochromatic. For $n=26$, the corresponding CNF is
\begin{equation}
	\begin{tabular}{l@{\hspace{0.7 cm}}l@{\hspace{1 cm}}l@{\hspace{1 cm}}}
\c{3 4 5} 	& \c{5 12 13 } & \c{6 8 10}  \\
\c{7 24 25} & \c{8 15 17 } & \c{9 12 15}  \\
\c{10 24 26}& \c{12 16 20} & \c{15 20 25} 
\end{tabular}
\end{equation}
together with the same clauses but with negated literals. Two generators are emitted by \tool{BreakID} for this formula, both of which simply swap the two colors. 
Now, suppose that the solver resolved $\c{3 4 -12 -13} $ from $\c{3 4 5}$ and $\c{-5 -12 -13}$. Symmetry does not help us detect that $\c{6 8 -24 -26}$ is a legitimate \eclause:  a resolvent of $\c{6 8 10},\c{-10 -24 -26}$. On the other hand we know that these two clauses are present in the formula because if $a,b,c$ form a Pythagorean triple, then so do $\alpha\cdot a,\alpha \cdot b, \alpha \cdot c$ for any $\alpha \in \mathbb{N}$, as long as they are within the bound $n$. Using this high-level information, we can safely add $\c{6 8 -24 -26}$. 
\qed
\end{example}
}

Since the subgraph isomorphism problem is NP-complete, we only focus on cases in which it can be indirectly inferred from analyzing the high-level structure of the original problem and controlling (or knowing) how it is encoded. Specifically, in such problems we derive a mapping between the literals, and adapt the solver to use this information in order to derive new \eclauses. Our implementation of this technique shows average overall improvement in terms of run-time.

To summarize, our contributions in this article are: 
\begin{enumerate} 
\item We show several problem domains in which this known principle can be exploited by the SAT solver for improving performance. So far it has only been used in bounded model checking and in neural network verification; 
\item We show how this technique is superior to, and can be seen as an extension of, dynamic symmetry; 
\item  We show how to modify the SAT-solver in order to implement this technique, and suggest several techniques for filtering \eclauses (i.e., decide which ones to keep, in light of possibly having too many of them) and deletion of such clauses;
\item We present experimental results that show certain performance improvements (around 50\% reduction in run time) due to this technique with domains in which it has not been used before.
\end{enumerate}
Although any paper that mentions an open mathematical problem such as Van der Warden numbers raises the expectation that it was able to solve it (i.e., find a new Van der Waerden number), this is not a result that can be found here: we only use it as one of several examples of problem domains in which the high-level structure can be used for improving run-time.

We continue in the next section by describing the method in detail. In Sec.~\ref{sec:examples} we will demonstrate how to apply it with several famous problems.

\section{Finding additional \eclauses} \label{sec:finding}

Let us recap. \textit{Symmetry} over $\varphi$ is a propositionally-consistent map $\sigma:lits(\varphi) \mapsto lits(\varphi)$ such that  $\sigma(\varphi) \equiv \varphi$. In this situation we can add symmetry-breaking constraints, and also use dynamic symmetry exploitation by adding \eclauses, but not both. 

\textit{Almost symmetries} refer to a situation where we have a formula $\varphi \equiv\varphi_1\cup\varphi_2$ and a propositionally-consistent map $\sigma:lits(\varphi_2) \mapsto lits(\varphi_2)$ such that $\sigma(\varphi_2) \equiv \varphi_2$.  Here we \emph{cannot} add symmetry-breaking constraints because of the $\varphi_1$ clauses, but we can still use dynamic symmetry exploitation by adding \eclauses that are based on $\varphi_2$. 

We now generalize almost symmetries as follows. Let
\begin{equation} \label{eq:main}
	\varphi \equiv\varphi_1\cup\varphi_2\cup\varphi_3\;,
\end{equation}
where $\varphi, \varphi_1,\ldots$ are sets of clauses, possibly overlapping. Let $\sigma:lits(\varphi_2) \mapsto lits(\varphi_3)$ be a literal map such that
\begin{equation}\label{eq:e23}
	\sigma(\varphi_2 )\equiv\varphi_3\;.
\end{equation}
Our central claim is:
\begin{proposition} \label{prop_main}
Let $c$ be a conflict clause that was learned from $\varphi_2$'s clauses, i.e., $\varphi_2\models c$. Then $\varphi$ and $\varphi \cup \sigma(c)$ have the same solutions.
\end{proposition}
\begin{proof}
Consider the resolution process by which $c$ was inferred from $\varphi_2$. The same resolution process can be applied to $\sigma(\varphi_2)$, and the result will be $\sigma(c)$.
Hence $\sigma(\varphi_2)\models\sigma(c)$, and because of (\ref{eq:e23}) we have  $\varphi_3\models\sigma(c)$.
Therefore, $\varphi \models\sigma(c)$ and we can add the \eclause $\sigma(c)$ to $\varphi$ without removing solutions.
\end{proof}
The following table summarizes the discussion so far. 
\begin{center}
	\begin{tabular}{lccc}
\toprule
		& Symmetry & Almost symmetry & \eclauses \\ \midrule
		Formula: & $\varphi$ & $ \varphi_1\cup\varphi_2$ & $ \varphi_1 \cup \varphi_2 \cup \varphi_3$ \\
		Requires: & $\sigma(\varphi)\equiv \varphi$ & $\sigma(\varphi_2)\equiv \varphi_2$ & $\sigma(\varphi_2)\equiv \varphi_3$ \\ \bottomrule
	\end{tabular}	
\end{center}

For a given formula $\varphi$, the question is how to define $\varphi_2,\varphi_3$ and the corresponding mapping $\sigma$ that satisfy (\ref{eq:e23}). As we will see in the next section, for certain types of formulas it can be done in such a way that \eclauses can be added in linear time. In fact it can be done in multiple ways, i.e., many such mappings exist, and we can use all of them.

\section{Examples} \label{sec:examples}
We will show here \Short{two} \Long{several} example problems that received attention in the SAT community in recent years, and in which \eclauses can be added efficiently \Short{: Van der Waerden numbers, and Boolean Pythagorean triples}. \Short{The long version of this article~\cite{IS21a} includes additional examples: Bounded model checking, SAT-based Planning, a combinatorial problem called `Sweep', and the anti-bandwidth problem.}

\subsection{Van der Waerden numbers (2 colors)}\label{sec:vdw}
We begin with the following definition:
\begin{definition}\label{def:vdw}
	The Van der Waerden number $W(j,k)$ is the smallest integer $n$ such that every 2-coloring of $1..n$ has a monochromatic arithmetic progression of length $j$ of color 1, or of length $k$ of color 2.
\end{definition}
For example, the following coloring proves that $W(3,3) > 8$, since there is no arithmetic progression of size 3 of either color:
\begin{center}
	 \scalebox{0.8}{\begin{picture}(0,0)%
\includegraphics{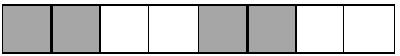}%
\end{picture}%
%
%
\setlength{\unitlength}{3947sp}%
\begingroup\makeatletter\ifx\SetFigFont\undefined%
\gdef\SetFigFont#1#2#3#4#5{%
  \reset@font\fontsize{#1}{#2pt}%
  \fontfamily{#3}\fontseries{#4}\fontshape{#5}%
  \selectfont}%
\fi\endgroup%
\begin{picture}(1905,260)(709,119)
\end{picture}%
}.
	 \end{center}

\noindent However, there is no such coloring for $n=9$, hence $W(3,3) = 9$. 

\Long{The following table shows the currently known Van der Waerden numbers when $j=k$ (concisely denoted by $w(k)$ in the table). As of $k > 6$ only lower-bounds are known. Even the case of $k = 6$ is far beyond the capability of modern SAT solvers. The value 1132 was proven by relying on mathematical argumentation in combination with a SAT solver for discharging millions of lemmas regarding smaller numbers~\cite{KP08}.

\begin{center}
\begin{tabular}{|c|c||c|c||c|c|} \hline
$k$ & $w(k)$ & $k$ & $w(k)$ & $k$ & $w(k)$ \\ \hline
3&9&	6&1,132	&	9& $>$ 41,265 \\ \hline
4&35&	7& $>$ 3,703 &10& $>$ 103,474   \\ \hline
5&178& 	8& $>$ 11,495 &11&$>$ 193,941  \\ \hline
\end{tabular}
\end{center}
}

There is relatively little symmetry in such formulas. An obvious one is the symmetry between the colors, when $j=k$.  Another type of symmetry is reversal (reading the sequence from the end). Reconsidering Example~\ref{ex:wd}, $\sigma_1,\sigma_2$ of (\ref{eq:br}) break these two symmetries. 

Given $j,k$ and $n$, encoding the decision problem whether $W(j,k) > n$ with CNF is simple.
Define $n$ variables $x_i$ for $1\leq i \leq n$, indicating whether location $i$ is assigned the color ‘1’. The constraints on the arithmetic progression are given by
\begin{equation} \label{eq:wd}
\begin{array}{l}
\{(x_i \vee x_{i+d} \vee \cdots \vee x_{i + (j-1)d}) \mid i\in [1, n - (j-1)d], d \geq 1\}  \\ 
\bigcup \\
\{(\bar{x}_i \vee \bar{x}_{i+d} \vee \cdots \vee \bar{x}_{i + (j-1)d}) \mid i\in[1, n - (k-1)d], d \geq 1\}\;,
\end{array}
\end{equation}
as was described, e.g., by Knuth in~\cite{K15}. From here on we will use integers as representatives of literals.

\begin{example} \label{ex:vdw}
Consider the case of $j = k = 3, n = 10$. When a variable $i$ is assigned true, it represents the decision to assign slot $i$ the color `1', and `0` otherwise. Then no 3 slots...
\begin{itemize}
\item ... with gap 1 are all ‘0’:   (1 2 3) (2 3 4) ... (8 9 10)
\item ... with gap 2 are all ‘0’:   (1 3 5) (2 4 6) ... (6 8 10)
\item ... with gap 3 are all ‘0’:   (1 4 7) (2 5 8) ... (4 7 10)
\item ... with gap 4 are all ‘0’:   (1 5 9) (2 6 10)
\end{itemize}
The same constraints, but with negated literals, are now added for the color ‘1’. For example, for gap 1, add (–1,–2, –3) ...  (-8, -9, -10), etc.
\hfill $\qed$
\end{example}

The clauses as defined in~(\ref{eq:wd}) have what we call a \emph{gliding symmetry}\footnote{Mathematicians use this term for describing a pattern that repeats itself by an operation of shifting in one dimension in space, e.g., $\spadesuit\ \spadesuit\ \spadesuit\ \spadesuit\ \ldots$}. This means that the same clause is replicated in the formula while shifting the variable index by a constant up to some bound, for example (1 2 3) is in $\varphi$, but also (2 3 4)...(8 9 10). Similarly (-1 -2 -3) is replicated with a negative constant. For a clause $c$, let $c^i_z$ denote the clause attained by taking $i$ steps towards zero, and similarly let $c^i_n$ denote the clause attained by taking $i$ steps away from zero, i.e., towards $n$ or $-n$.
For example $\c{3 4 5}^1_z = \c{2 3 4}$ and $\c{1 2 3}^1_n = \c{2 3 4}$. As another example, this time focusing on the negative constraints, $\c{-1 -3 -5}^1_n = \c{-3 -5 -7}^1_z = \c{-2 -4 -6}$.

For each clause $c \in \varphi$, we save the \emph{gliding bounds} $[i,j]$, where $i,j$ are the maximal integers such that $c^i_z, c^j_n \in \varphi$. For example, for the clause $c = \c{2 3 4}$ of Example~\ref{ex:vdw}, we save the pair [1, 6], because we can `glide' by up to one step towards zero and by up to six steps towards $n = 10$ (giving us, respectively, (1 2 3) and (8 9 10)). As another example, the pair for the clause (-4 -5 -6) is [3,4], because we can glide by up to three steps towards zero, and by up to four steps towards $-n = -10$. Denote by $c.z$ and $c.n$ the two bounds of a clause $c$, corresponding to $i,j$ above, respectively. 

So far we only considered the original clauses of the problem. We now consider the question of what are the bounds for the learned clauses. Let $c_1,\ldots,c_m$ be the antecedent clauses of a new learned clause $c$. We compute the gliding bounds of $c$ as follows:
\begin{equation}
\label{eq:cz}
c.z  =  \min(c_1.z,\ldots,c_m.z) \qquad
c.n  =  \min(c_1.n,\ldots,c_m.n)\;.
\end{equation}
The rational of (\ref{eq:cz}) is that we can only glide $c$ towards zero (or away from zero) as much as we can glide all of its antecedents towards zero (or away from zero).

Given the gliding bounds of each clause, it is easy to use Proposition~\ref{prop_main} for learning new \eclauses.
Using the terminology of that proposition, the antecedents of $c$ form $\varphi_2$, and $\sigma$ is a mapping that applies `gliding' to them. Each amount of gliding is a separate mapping $\sigma$. The gliding bounds tell us the amount by which gliding each clause results in a clause that is still in $\varphi$ -- those new clauses are $\varphi_3$ in the proposition. In other words, those bounds define the mappings that we can use for deriving new \eclauses.

\begin{example}\label{ex:eclause}
Suppose $\varphi$ includes the following clauses and respective bounds:
\begin{equation}\label{eq:ex12}
  \c{3 6 10}  [2,0] \quad
  \c{-7 -5 -3} [2,2] \quad
  \c{-7 -6 -5} [4,2]
\end{equation}
from which the solver inferred via resolution the clause $c = \c{-7 -5 10}$. With (\ref{eq:cz}) we compute the gliding bounds [2,0] for $c$. This means that we have two mappings:
\begin{itemize}
  \item $\sigma_1$ maps each positive literal $l$ to $l-1$ and negative literal $-l$ to $-l+1$
  \item $\sigma_2$ maps each positive literal $l$ to $l-2$ and negative literal $-l$ to $-l+2$,
\end{itemize}
i.e., a glide by one and two towards 0.
So we add the \eclauses $\sigma_1(c) = \c{-6 -4 9}$ and $\sigma_2(c) = \c{-5 -3 8}$. Indeed, if we apply $\sigma_1$ to the clauses in (\ref{eq:ex12}), we get three clauses in $\varphi$, from which we can infer $\sigma_1(c)$:
\[
\begin{array}{ll}
\sigma_1\c{3 6 10} = \c{2 5 9} &
  \sigma_1\c{-7 -5 -3}  = \c{-6 -4 -2} \\
  \sigma_1\c{-7 -6 -5}  = \c{-6 -5 -4}  & 
\end{array} 
\]
\hfill $\qed$
\end{example}

Finally, we should compute the gliding bounds of the \eclauses themselves, because they may participate in further learning. For this, we shift the bounds of the conflict clause by the same amount as dictated by the mapping $\sigma$, while recalling that any step towards zero is a step away from $n$ (or $-n$ if it is a negative literal), and vice a versa.
\begin{example}
Reconsider $c$ of Example~\ref{ex:eclause}. Its bounds are [2,0]. We computed
$\sigma_1(c)$ by gliding $c$ towards zero by 1. Hence the bounds of $\sigma_1(c)$ are [2$-$1,0$+$1] = [1,1].
\hfill $\qed$
\end{example}

\Long{
\subsection{A sweep problem}
Given natural numbers $m,n,r,k$, the `sweep' problem is to decide whether there is a \{0,1\} matrix $X$ of size $m\times n$ such that at least $r$ entries are ‘1’, and none of its submatrices has a diagonal (from northwest to southeast) with more than $k$ ‘1’ entries. Formally,
\[ \begin{array}{l@{\hspace{1.5 cm}}}
sweep(X) = max_{
	\begin{array}{l}
		i_1 \leq \cdots \leq i_t \leq m \\
		j_1 \leq \cdots \leq j_t \leq n 
		\end{array}}
	\Sigma_{l=1}^t x_{i_l,j_l}
	\end{array}
\]
and we require $sweep(X) \leq k$.
For example, the following matrix is a solution for $m=4,n=3,r=6,k=2$.

\[ 
{\small	\left(
	\begin{tabular}{ccc}
		1 & 0 & 1 \\
		1 & 0 & 0 \\
		1 & 1 & 0 \\
		0 & 1 & 0 \\
	\end{tabular}
	\right)
}
\]

An encoding suggested by Knuth in~\cite{K15} is based on variables $x_{i,j}$ indicating whether entry $i,j$ is ‘1’ for $1\leq i\leq m,1\leq j \leq n$, and $s_{i,j}^t$, which is true if the sweep value of the submatrix $[(1,1),(i,j)]$ of $X$ is at least $t$, for $i,j$ in the same range and $1\leq t \leq \min(m,n)$. 

We will not describe here the full set of constraints. We will only show that some of them (in fact, almost all) have gliding symmetry, and hence enable \eclauses. Recall that 
constraints that do not have this property fit into the $\varphi_1$ part of (\ref{eq:main}), and according to Proposition~\ref{prop_main} we can still add \eclauses, as long as the base clause was inferred from $\varphi_2$. 

First, let us focus on these two constraints: 
\begin{equation}\label{eq:sweep}
	s_{i,j}^t  \rightarrow  s_{i+1,j+1}^t \quad \qquad
	s_{i,j}^t \wedge x_{i+1,j+1}  \rightarrow  s_{i+1,j+1}^{t+1} 
\end{equation}
for $1 \leq i < m, 1 \leq j < n$. Writing these implications as clauses reveal the gliding symmetry along the diagonals, e.g., for the first of those, the clauses are $(\neg s_{1,1}^1 \vee s_{2,2}^1), (\neg s_{2,2}^1 \vee s_{3,3}^1), \ldots$. 
Two other constraints for this problem that have a gliding symmetry are 
	$s_{i-1,j}^t  \rightarrow s_{i,j}^t$ and 
	$s_{i,j-1}^t  \rightarrow s_{i,j}^t$, 
with similar ranges of $i$ and $j$.  
Finally, this problem contains a cardinality constraint, forcing at least $r$ `1' entries.
Using a sequential counter to encode this constraint, as suggested by Sinz in~\cite{S05},
gives us another set of constraints, that most of them (specifically, all except a clause at the beginning and at the end of the counter) also have a gliding symmetry. This means that we can add them to $\varphi_2$ and therefor increase the number of \eclauses.

Like in the case of Van der Waerden formulas, here too we need to compute the gliding bounds. The principle is the same, hence we won't detail it here.  
}
\Long{
\subsection{The anti-bandwidth problem}
Given a graph with nodes $1..n$ and a number $1 \leq k \leq n$, the 
anti-bandwidth problem (ABP) is to decide whether there is a mapping (`labeling') of each node to a number in the range $0..n$ such that each label is used at most once, and the smallest difference between labels of neighboring nodes is at least $k$. This graph-labeling problem has various applications such as scheduling and radio frequency assignment, and was also used recently in~\cite{FSBP20} as a case study for a new type of SAT encoding, called `staircase encoding'. Please see~\cite{FSBP20} for a survey of ABP-related research.  

A natural way to encode this problem into SAT, is to define the set of variables $\{x_i^j\mid 1\leq i \leq n, 1\leq j \leq n\}$, where $x_i^j$ is true if and only if node $i$ is labeled with label $j$. There are constraints forcing each node to be labeled with a single label, and each label being assigned at most once. For our purpose of detecting a potential for \eclauses beyond symmetry, we focus however on the constraint that forces the gap between two neighboring nodes to be larger or equal to $k$. For a given edge $(i,i')$, the constraint is that up to one variable from the set \[\{x_i^{l + \gamma},x_{i'}^{l + \gamma} \mid 1 \leq l \leq n-k, 0 \leq \gamma \leq k\} \]
can be true. For example, if $k=2, n=5$, then we have (in a `at most one' constraint form)

\begin{equation}\label{eq:abp}
\begin{array}{lll}
x_i^1 + x_{i'}^1 + x_i^2 + x_{i'}^2 + x_i^3 + x_{i'}^3 & \leq & 1 \\
x_i^2 + x_{i'}^2 + x_i^3 + x_{i'}^3 + x_i^4 + x_{i'}^4 & \leq & 1 \\
x_i^3 + x_{i'}^3 + x_i^4 + x_{i'}^4 + x_i^5 + x_{i'}^5 & \leq & 1
\end{array}
\end{equation}
The gliding symmetry in~(\ref{eq:abp}) is apparent. One thing that is not a priori clear is whether a solver can learn something from these clauses alone that is not subsumed by existing clauses. This depends on the exact encoding of these at-most-one constraints. If it reduces to adding a constraint for each pair of variables in the constraint, forcing at least one of them to be false, then no further learning is possible, and hence \eclauses become irrelevant. 
}
\Long{
\subsection{Bounded model checking and planning} \label{sec:bmc}
Bounded model checking (BMC)~\cite{BCCZ99} and Planning~\cite{KS92} are old problems that are routinely solved with SAT solvers, and have essentially the same structure. The former attempts to find an error in an open synchronous system (i.e., a system that is controlled by inputs in each step) in its $k$'s step, and the latter attempts to find a $k$-long set of steps that lead from a given initial state to a goal state. In both cases $k$ is increased if the formula turns out to be unsatisfiable, until it becomes satisfiable or intractable in terms of computational resources.

Let $S$ be a set of Boolean variables that encode the state of the system, and $S_i, 0 \leq i \leq k$ be copies of those variables, representing the state after $i$ steps, $S_0$ being the initial state. Let $I$ denote a predicate over $S_0$ that defines the initial state, $T(S_i,S_{i+1})$ for $0\leq i \leq k$ be the transition relation, and $P$ the goal state (in the case of BMC, $P$ is the negation of the property that is being checked). The SAT encoding in both problems has the following structure:
\begin{equation}\label{eq:bmc}
  I(S_0) \wedge \bigwedge_{i\in[0,k-1]}T(S_i,S_{i+1}) \wedge P(S_k).
\end{equation}
The projection of an assignment that satisfies (\ref{eq:bmc}) to the inputs, gives us a trace to a bug in case of BMC, and a plan in case of a planning problem. The regular structure of this equation can be exploited for detecting isomorphic subgraphs. In fact, for the case of BMC, this was already done many years ago in~\cite{S00}, although without the analysis via isomorphism as done here; rather it showed a direct technique for adding clauses -- here called \eclauses -- based on the structure of (\ref{eq:bmc}).

The regularity of the big conjunction in (\ref{eq:bmc}) represents a gliding symmetry, similarly to the case of Van der Waerden numbers that we presented before. Specifically, $T(S_i,S_{i+1})$ is represented with clauses that are isomorphic to those representing $T(S_j,S_{j+1})$ for $0\leq j \leq k-1$, all of which are in $\varphi$. Hence here we also need to keep gliding bounds for each clause, and update them following the same principles that were described in Sec.~\ref{sec:vdw}. Here, however, the formula contains asymmetric parts, namely $I(S_0)$ and $P(S_k)$. We assign them the gliding bounds [0,0], which guarantees that whenever they are antecedents of a conflict clause, that clause will not be used for deriving \eclauses, neither at the point that it was learned, nor in later steps, if this clause participates in resolving new clauses.
} 

\subsection{Boolean Pythagorean triples}\label{sec:pyth}
We conclude with an example that shows that \eclauses are not necessarily tied to gliding symmetry. 

Three positive integers $a,b,c$ are called a Pythagorean triple if they satisfy $a^2 + b^2 = c^2$.
The challenge is:  
\begin{definition} \label{def:pyth}
For a given $n \in N$, can $1..N$ be separated into two sets, such that no set contains a Pythagorean triple?
\end{definition}
As an example, for $n=17$ if we choose the subset of integers that is here marked with an underline: 1 2 3 4 \underline{5} 6 7 \underline{8} \underline{9} 10 11 12 13 14 15 16 17, it proves that for $n=17$ the answer is yes.

The general question of whether there exists an $n$ for which the answer is negative was open for many years. The celebrated result of Heule et al.~\cite{HKM16} a few years ago proved, with the help of a SAT solver, that the answer is positive.

The encoding of the problem in Def.~\ref{def:pyth} with CNF is very simple: define $n$ variables, where the Boolean values in the satisfying assignment separate the values naturally to the two requested sets. For example, the encoding for $n=17$ is
\[
\begin{array}{ll@{\hspace{0.7 cm}}ll}
  \c{3 4 5} & \c{-3 -4 -5} &
  \c{5 12 13} & \c{-5 -12 -13}  \\
  \c{6 8 10} & \c{-6 -8 -10} &
  \c{8 15 17} & \c{-8 -15 -17} \\
  \c{9 12 15} & \c{-9 -12 -15} 
\end{array}
\]
Denote by $\varphi_n$ this formula for a given $n$.
In the discussion that follows we will overload the multiplication and division signs, '$\cdot$' and '$/$' to operate on clauses and sets of clauses: the operation is simply applied to each of the literals. For example, $2\cdot \c{3 4 5} = \c{6 8 10}$ and $\c{6 8 10}/2 = \c{3 4 5}$.

We begin with two simple observations:
\begin{observation}\label{obs:1}
Pythagorean triples are closed under multiplication:
\[\forall a,b,c,i\in N.\ a^2 + b^2 = c^2 \Rightarrow (a\cdot i)^2 + (b\cdot i)^2 = (c\cdot i)^2\;.\]
\end{observation}
\begin{observation} \label{obs:2}
Let $|_d$ denote `divisible by $d$'. When applied to a set of numbers, then it means that all the set's members are divisible by $d$. Then for all $n$,
\begin{equation}
\c{a b c} \in \varphi_n \wedge \c{a b c}|_d \Rightarrow \frac{\c{a b c}}{d} \in \varphi_n\;.
\end{equation}

\end{observation}
The second observation is simply the other side of the first one (dividing rather than multiplying), but it also states that the divided clause must be in $\varphi_n$. For example, if $n=80$ then $\c{30 72 78} \in \varphi_{80}$, which implies that also (30 72 78)/2 = (15 36 39) $ \in \varphi_{80}$.

For each clause $c$, we define recursively
\begin{equation}\label{eq:gcd}
c.gcd = \left\{\begin{array}{ll}
gcd(\{l \mid l \in c\}) & $c$ \mbox{ is original} \\
gcd (\{c_i.gcd \mid c_i \in S\}) & $c$ \mbox{ is inferred from } \\
& \mbox{a clause set } $S$
\end{array}\right.
\end{equation}
where $gcd()$ is the greatest common divider function. Observe that if $c$ is original, then $c.gcd$ is the greatest common divider of its own variables, and otherwise of
the variables in the core of original clauses that derived it, which we will denote by $core(c)$. This recursive definition gives us an immediate method to implement it in a SAT solver: the base case corresponds to the original clauses, and the step to the learning that is done during conflict analysis.

Given a conflict clause $c$, we can see that for $i\in[1,bound(n)]$ ($bound(n)$ will be defined shortly), we have
\begin{equation}\label{eq:fact}
i \cdot \frac{core(c)}{c.gcd}\subseteq \varphi_n\;.
\end{equation}
This is a direct result of the two observations above: From Observation~\ref{obs:2} we know that $\frac{core(c)}{c.gcd}\subseteq \varphi_n$, and from Observation~\ref{obs:1} we know that any multiplication of this clause is a Pythagorean triple. Whether it is part of $\varphi_n$ depends on the value of $i$, which brings us to the problem of computing $bound(n)$. To compute it, we need to know the largest variable that participates in $core(c)$. For each clause $c$, we define recursively
\[
c.maxvar = \left\{
\begin{array}{ll}
\max(\{l \mid l \in c\}) & $c$ \mbox{ is original} \\
\max (\{c_i.maxvar \mid c_i \in S\}) & $c$ \mbox{ is inferred }  \\
& \mbox{from a set } $S$
\end{array}\right.
\]
Hence, for each clause $c$, $c.maxvar$ denotes the largest variable that appears in $core(c)$.
In (\ref{eq:fact}) we considered clauses $i \cdot \frac{core(c)}{c.gcd}$. For these clauses to be part of $\varphi_n$, the following relation should hold:
\[i \cdot \frac{c.maxvar}{c.gcd} \leq n\;.\]
Isolating $i$ gives us the bound:
$bound(n) = \frac{n\cdot c.gcd}{c.maxvar}.$ 
Finally, observe the implication of (\ref{eq:fact}): since $i \cdot \frac{core(c)}{c.gcd}\subseteq \varphi_n$, then
\begin{equation}
\varphi_n \models i \cdot \frac{c}{c.gcd}, \mbox{ for } i\in[1,bound(n)]\;.
\end{equation}
This means that $i \cdot \frac{c}{c.gcd}$ can be added safely as \eclauses to $\varphi_n$, without removing solutions. In other words, using the terminology of Sec.~\ref{sec:finding}, each $i\in[1,bound(n)]$ defines us a separate mapping for a conflict clause $c$:
\begin{equation}
\sigma_i(c) = i \cdot \frac{c}{c.gcd}\;.
\end{equation}

\section{Implementation details}\label{sec:impl}
Recall that according to~(\ref{eq:main}) the formula may contain a non-empty set of clauses $\varphi_1$, that cannot participate in generating \eclauses. In our implementation we mark those clauses at the beginning (such clauses are expected to be given in a separate input file), and then also each learned clause that one of its antecedents is marked that way. For simplicity let us call these clauses \textit{non-symmetric} and the rest \textit{symmetric}. 

\Long{We experimented with two methods of keeping track of this information. The first is via activation literals and assumptions. One simply adds such a literal to each non-symmetric clause in the input formula, and activate those clauses by sending its negation as an assumption to the solver. These activation literals are \emph{pure} and hence never disappear during resolution. As a result, whenever they are used in order to derive a new clause, the activation literal is part of that new clause as well. Hence, upon learning a new conflict clause, the presence of an activation literal implies that we cannot use it for deriving \eclauses. While this is a simple method to implement, it has a rather significant performance penalty: each learned clause now includes an additional literal if the non-symmetric part of the formula is part of its premise. 
An alternative method that we implemented is based on altering the solver itself to keep track of dependencies. 
}
\Short{To keep track of these dependencies, we altered the solver.}
This is a non trivial task because logical dependency between clauses is created in many different parts of a modern solver. In particular, our implementation is based on \tool{Maple\_LCM\_Dist\_ChronoBT}~\cite{NR18} (we will abbreviate its name to \tool{chrono} from hereon), the winner of the SAT competition in 2018, which in itself is built on top of multiple generations of optimizations that were added to it over the years, all the way up to \tool{Minisat-2.2}~\cite{minisat}.  
In particular, dependency is created during conflict analysis in the process of learning a new clause, but also during clause minimization, binary-resolution minimization, learnt-clause simplifications, var elimination and propagation at decision level 0\footnote{These are implemented in the following functions in \tool{Chrono}: analyze, LitRedundant, binResMinimize, simplifyLearnt, eliminateVar, propagate}. 
We maintain a single bit in the header of each clause that determines whether it is symmetric or not. Since \tool{Chrono}, like all \tool{Minisat}-based solvers, do not maintain unit clauses, we maintain a separate list of variables that their value is determined at level 0 based on non-symmetric clauses. 

Next, we need to maintain problem-specific information that is necessary for deriving \eclauses. 
For example, for Van der Waerden formulas\Long{, bounded model-checking and planning problems} -- see Secs.~\ref{sec:vdw} \Long{and \ref{sec:bmc}} --  we need to keep for each clause its gliding bounds. For the Boolean Pythagorean triples problem -- see Sec.~\ref{sec:pyth} -- we maintain the greatest common divider (gcd) of the literals in the clause and all clauses that participated in deriving it, and the max variable in those clauses. As in the case of the symmetry bit described above, here too we need to update this information in every location in which dependency is created. 

Our implementation accumulates \eclauses and then adds them to the clause database at the nearest restart. This is a different strategy than the ones mentioned in the introduction in the context of symmetric explanation learning~\cite{DBB17} and dynamic symmetry handling~\cite{DBB17,TD19}, where such clauses are added during BCP, hence affecting the current search branch (we implemented both, and the results are rather similar, with a small advantage to the technique described here). 
To reduce side-effects, upon adding a new \eclause we do \emph{not} increase the counter of conflict clauses, since that counter affects various other heuristics, such as the frequency of applying simplifications and clause deletion. 

The above-mentioned prior works describe various filtering methods: adding clauses only if they conflict the current state or lead to further propagation, or, in the case of~\cite{TD19}, if the conflict clause itself has a low LBD. \Short{Several filtering and deletion strategies that we experimented with are described in the long version of this article~\cite{IS21a}. Briefly, the ones we settled on as best in our experiments are (1) add an \eclause only if up to 3 literals are not false under the current partial assignment, and (2) do not add \eclauses larger than 20. As for deletion strategies, we (1) gave a separate initial activity score of 0.8 for \eclauses and (2) set the deletion ratio to 0.8, i.e., a more aggressive deletion comparing to the default of 0.5. We left this deletion ratio also for the experiments without \eclauses, for a fair comparison.}

\Long{
We will now describe the filtering and deletion strategies that we implemented. Recall that \eclauses are added at restarts.
\subsubsection{Filtering strategies} 
We implemented the following filtering strategies: 
\begin{itemize}
	\item[F1] Add an \eclause only if up to $x$ literals, $x$ being a user-defined value,  are not false under the current partial assignment. When $x=0$ it means that we only add clauses that contradict the current partial assignment, and when $x=1$ we allow in addition clauses that are asserting under the current partial assignment.  
	\item[F2] Add an \eclause only if its \emph{partial} LBD is smaller than a user-defined value. The LBD is `partial' in the sense that the \eclause may contain literals that are not assigned under the current partial assignment, so we do not count them.
	\item[F3] Add an \eclause only ifs size is under a user-defined value. 

	\item[F4] An explicit bound on the number of added \eclauses. 
	\item[F5] A bound on the number of \eclauses that are being examined with respect to F1,F2 above, to reach the bound from F4.
\end{itemize}
The combination that worked best on average in our experiments is F1 with $x = 3$, and F3 with a bound of 20.

\subsubsection{Deletion strategies}
We implemented the following deletion strategies: 
\begin{itemize}
	\item[D1] A separate initial activity score for \eclauses. 
	\item[D2] A control of the initial learnt-clause list into which the \eclauses are inserted. Modern Minisat-based solvers add each learnt clause to one of three such lists: `Local', `Tier2' and `Core', based on their predicted importance. `Local' is the list of the least important clauses (in \tool{Chrono} this list is populated with references to learned clauses with LBD $> 6$). During periodic clause database reductions, half of the clauses in this list are removed. If clauses are sufficiently active or their LBD becomes smaller, they may move to `Tier2' or 'Core', which are protected from being reduced. Clauses can also move back from `Tier2' to `Local' if they become less active.
	\item[D3] The ratio of clauses to be deleted from the `Local' list. Default is 0.5. 
\end{itemize}
The combination that worked best on average in our experiments is D1 with a score of 0.8 (comparing to the weight 1 given to normal conflict clauses), D2 with the list `Local', and D3 with the ratio 0.8 (we left this deletion ratio also for the experiments without \eclauses, for a fair comparison). 
}

\section{Results}
We implemented this method for Van der Waerden numbers and Boolean Pythagorean triples. Since there is no standard benchmark sets for these problems, we generated instances, and took all of those that can be solved with at least one configuration in less than 30  min., and with at least one configuration in more than 1 min. For the Van der Waerden problems, this resulted in 30 benchmarks (16 unsat, 14 sat). The benchmarks, full tables of results, and the implementation are available from~\cite{url21}. We used the \tool{Hbench} benchmarking system~\cite{hbench} to conduct the experiments and data collection.

In the results tables below, timed-out benchmarks contribute the values they had at the timeout point to the various columns, other than the \textbf{par-2} column, where the timeout is added twice, to be consistent with the ranking method of the SAT competitions. Our goal was mostly to measure the number of \eclauses that can be found based on isomorphic subgraphs, beyond what can be found with dynamic symmetry exploitation.  We have evidence from multiple previous works, e.g.,~\cite{DBB17,TD19,S00} (see Sec.~\ref{sec:intro}), that such clauses can help in reducing the run time. Our results below show not only that indeed many more such clauses can be generated, but also that when combined with the right filtering and deletion methods, it reduces the run time on average. 

The results for the Van der Waerden problems are summarized in Table \ref{tab:vdw}, sorted by performance. The `-waerden' flag indicates that \eclauses are added as described in Sec.~\ref{sec:vdw}. The `-dyn-sym-exploit' flag indicates that \eclauses based on dynamic symmetry exploitation were added. `native' means that the solver was run in its default configuration \Long{(other than D3 -- see deletion strategies in Sec.~\ref{sec:impl})}\Short{other than the deletion ratio -- see Sec.~\ref{sec:impl}}. `static-sym-breaking' indicates that we solved the formula with static symmetry-breaking constraints, as provided by \tool{BreakID}, while the solver is in the same configuration as `native'. For these benchmarks static symmetry breaking turns out to be better than  dynamic symmetry exploitation, based on the same data (even when considering the unsat cases on their own). 
\Long{When considering unsat formulas only (not shown), static symmetry breaking is the best of all configurations.} 

\begin{table*}
	\begin{center}
	\begin{tabular}{lcccccccc} \toprule
\textbf{Configuration}& \textbf{Timed-}	& \textbf{Time}	& \textbf{Time}	& \textbf{Conflicts} & \textbf{\eclauses} & \textbf{Over-} & \textbf{Active}& \textbf{Active}\\ 
&\textbf{out}&&\textbf{(par-2)}&&&\textbf{head}& -E-& -C- \\ \midrule
-waerden	& 0			& 111.2		& 111.2	& 1,079,719		& 30568 	& 6  	& 0.017	& 0.015\\ 
-static-sym-breaking& 1	& 149.8		& 211.2	& 2110472
& 0			& 0		& 	&  \\ 
(native)	& 1			& 190.4		& 251.7	& 2,112,666		& 0 		& 0 	& 		& 0.011\\ 
-waerden -dyn-sym-exploit  & 2	& 198.5		& 317.7	& 1,963,104		& 50618 	& 10 	& 0.014 & 0.008 \\  
-dyn-sym-exploit & 3	& 233.2		& 418.6	& 2,477,840		& 6,729 	& 3 	& 0.011		& 0.008 \\ \midrule
-waerden	& 6			& 476.5		& 841.9	& 750,453		& 16,556,216& 29 	& 0.013 & 0.013	 \\ 
-dyn-sym-exploit & 1	& 119.8 	& 181.3 & 1,248,573		& 1,290,402 & 13	& 0.008 & 0.011   \\ \bottomrule
\end{tabular}
\caption{Average results for the Van der Waerden problem, over 30 benchmarks. Time is in seconds. The last two rows refer to runs without any filtering of the \eclauses\Long{, which turns out to be better for dynamic symmetry exploitation}.
} \label{tab:vdw}
\end{center}
\end{table*}

On average each conflict clause learned while solving these benchmarks results in over 20 \eclauses with the -waerden flag (this clearly depends on the value of $n$), and less than 1 with -dyn-sym-exploit. The latter is expected, since \tool{BreakID} generates a single generator for these benchmarks (see text after Def.~\ref{def:vdw}). The top part of the table does not reflect these numbers, however, because it refers to runs in which we applied aggressive filtering as mentioned before. With these filters, the number of \eclauses added is typically less than 5\% of the total number of clauses. Hence the potential for \eclauses is large, and perhaps future research into filtering techniques will be able to exploit this unused potential. The overhead of generating the \eclauses is marginal (the `Overhead' column). The overhead of running \tool{BreakID}, a necessary step for applying both -dynamic-symmetry and -symmetry-breaking, was a few seconds and not included in the `Time' column. 

We can see a run-time reduction of 42\% comparing to a native run for the case of Van der Waerden formulas, and of 55\% for the case of Pythagorean triples. In both cases the technique as described in \ref{sec:vdw} is better than adding \eclauses based on data derived from static symmetry, and better than combining these two sources of data. Cactus plots for both families appear in Figs.~\ref{fig:vdw} and ~\ref{fig:pyth}.

We also checked how active the \eclauses are in deriving new clauses. For this measure we define as \eclauses, recursively, the set of clauses that we add directly and the clauses that were learned based on at least one \eclause premise.
Activity of clauses is updated in the solver in the usual way, based on their participation in deriving other clauses. Since clause deletion is based on this activity, the ratio between the average number of `live' clauses (i.e., that were not deleted) and the total number of learned clauses is an indication of how active they are. This ratio for \eclauses and normal conflict clauses appear in the last two columns of the table. It is surprising to see that the \eclauses are more active, especially \Long{in the context of deletion strategy D1, namely the fact that}\Short{since } we initiate the activity score of \eclauses with a lower value in comparison to the value given to conflict clauses. 

For the Boolean Pythagorean triples problem, we generated 21 satisfiable instances (the first unsatisfiable instance takes weeks to solve --- see~\cite{HKM16}) with the same selection criteria as described above. The results appear in Table~\ref{tab:pyth}, also in ascending performance order. Here the native solver turns out to be improved-upon in each of the configurations, including static symmetry breaking.

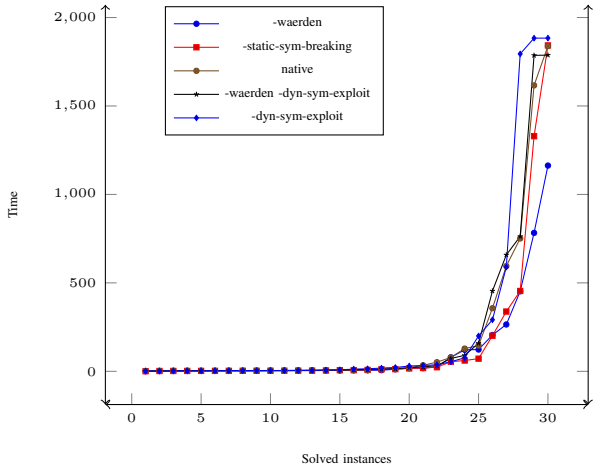
\begin{figure}
	\usetikzlibrary{backgrounds}
\tikzset{
	extra padding/.style={
		show background rectangle,
		inner frame sep=#1,
		background rectangle/.style={
			draw=none
		}
	},
	extra padding/.default=0.1cm,
}
\pgfplotsset{width=8cm, 
	every axis/.append style={                    
		axis line style={<->}, 
		label style={font=\tiny},
		tick label style={font=\tiny}
}}
%
\begin{tikzpicture}[extra padding]
	\begin{axis}[
		title={},
		xlabel={Solved instances},
		ylabel={Time},
		axis x line*=none,
		axis y line=box, 
	    legend style={at={(0.35,+5.3cm)}, 
			anchor=north,legend columns=1, font = \tiny},
		nodes near coords,
		point meta=explicit symbolic,
		]
		\addplot+[mark size=1pt] coordinates {  
			(1, 0.88) [ ]
			(2, 1.94) [ ]
			(3, 2.84) [ ]
			(4, 3.03) [ ]
			(5, 3.09) [ ]
			(6, 3.23) [ ]
			(7, 3.49) [ ]
			(8, 3.55) [ ]
			(9, 3.60) [ ]
			(10, 4.17) [ ]
			(11, 4.45) [ ]
			(12, 4.50) [ ]
			(13, 4.70) [ ]
			(14, 5.04) [ ]
			(15, 5.12) [ ]
			(16, 5.45) [ ]
			(17, 5.59) [ ]
			(18, 6.41) [ ]
			(19, 11.64) [ ]
			(20, 17.81) [ ]
			(21, 18.10) [ ]
			(22, 24.88) [ ]
			(23, 79.17) [ ]
			(24, 121.52) [ ]
			(25, 123.13) [ ]
			(26, 205.09) [ ]
			(27, 264.72) [ ]
			(28, 452.42) [ ]
			(29, 782.55) [ ]
			(30, 1163.11) []
		};
		\addplot+[mark size=1pt] coordinates {  (1, 0.53) [ ]
			(2, 1.76) [ ]
			(3, 2.02) [ ]
			(4, 2.15) [ ]
			(5, 2.46) [ ]
			(6, 2.51) [ ]
			(7, 2.63) [ ]
			(8, 3.20) [ ]
			(9, 3.34) [ ]
			(10, 3.43) [ ]
			(11, 3.56) [ ]
			(12, 3.86) [ ]
			(13, 4.00) [ ]
			(14, 4.58) [ ]
			(15, 6.25) [ ]
			(16, 6.58) [ ]
			(17, 8.14) [ ]
			(18, 11.38) [ ]
			(19, 12.28) [ ]
			(20, 16.42) [ ]
			(21, 18.87) [ ]
			(22, 23.14) [ ]
			(23, 54.38) [ ]
			(24, 61.04) [ ]
			(25, 71.70) [ ]
			(26, 201.25) [ ]
			(27, 337.91) [ ]
			(28, 454.28) [ ]
			(29, 1329.13) [ ]
			(30, 1842.20) []
		};
		\addplot+[mark size=1pt] coordinates {  (1, 2.24) [ ]
			(2, 2.49) [ ]
			(3, 2.59) [ ]
			(4, 2.74) [ ]
			(5, 2.80) [ ]
			(6, 2.86) [ ]
			(7, 3.23) [ ]
			(8, 3.32) [ ]
			(9, 3.36) [ ]
			(10, 4.16) [ ]
			(11, 4.24) [ ]
			(12, 5.26) [ ]
			(13, 5.27) [ ]
			(14, 5.29) [ ]
			(15, 5.37) [ ]
			(16, 7.55) [ ]
			(17, 7.78) [ ]
			(18, 11.47) [ ]
			(19, 15.58) [ ]
			(20, 19.41) [ ]
			(21, 34.35) [ ]
			(22, 51.62) [ ]
			(23, 78.26) [ ]
			(24, 128.47) [ ]
			(25, 145.79) [ ]
			(26, 356.64) [ ]
			(27, 594.97) [ ]
			(28, 751.00) [ ]
			(29, 1616.38) [ ]
			(30, 1838.83) []
		};
		\addplot+[mark size=1pt] coordinates {  (1, 1.11) [ ]
			(2, 2.18) [ ]
			(3, 2.89) [ ]
			(4, 3.28) [ ]
			(5, 3.37) [ ]
			(6, 3.56) [ ]
			(7, 3.70) [ ]
			(8, 3.82) [ ]
			(9, 3.90) [ ]
			(10, 4.26) [ ]
			(11, 4.73) [ ]
			(12, 5.52) [ ]
			(13, 5.70) [ ]
			(14, 7.28) [ ]
			(15, 7.82) [ ]
			(16, 8.96) [ ]
			(17, 10.42) [ ]
			(18, 11.75) [ ]
			(19, 15.08) [ ]
			(20, 20.26) [ ]
			(21, 25.37) [ ]
			(22, 31.72) [ ]
			(23, 72.52) [ ]
			(24, 90.40) [ ]
			(25, 158.51) [ ]
			(26, 454.06) [ ]
			(27, 658.05) [ ]
			(28, 762.64) [ ]
			(29, 1786.21) [ ]
			(30, 1787.35) []
		};
		
		\addplot+[mark size=1pt] coordinates {  (1, 1.72) [ ]
			(2, 1.85) [ ]
			(3, 2.09) [ ]
			(4, 2.60) [ ]
			(5, 2.74) [ ]
			(6, 3.08) [ ]
			(7, 3.66) [ ]
			(8, 3.73) [ ]
			(9, 3.80) [ ]
			(10, 4.15) [ ]
			(11, 4.21) [ ]
			(12, 4.33) [ ]
			(13, 5.68) [ ]
			(14, 6.52) [ ]
			(15, 8.28) [ ]
			(16, 12.47) [ ]
			(17, 14.33) [ ]
			(18, 18.19) [ ]
			(19, 21.59) [ ]
			(20, 29.81) [ ]
			(21, 32.85) [ ]
			(22, 36.74) [ ]
			(23, 53.99) [ ]
			(24, 76.50) [ ]
			(25, 199.87) [ ]
			(26, 291.37) [ ]
			(27, 589.39) [ ]
			(28, 1794.39) [ ]
			(29, 1883.45) [ ]
			(30, 1883.67) []
		};
		
		\legend{ -waerden, -static-sym-breaking, native,  -waerden -dyn-sym-exploit, -dyn-sym-exploit};
	\end{axis}
\end{tikzpicture}\caption{Results for the Van der Waerden benchmarks.}
	\label{fig:vdw}
\end{figure}
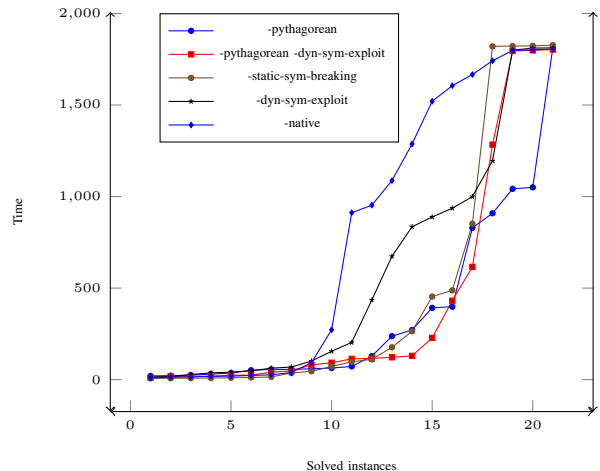
\begin{figure}
	\usetikzlibrary{backgrounds}
\tikzset{
    extra padding/.style={
        show background rectangle,
        inner frame sep=#1,
        background rectangle/.style={
            draw=none
        }
    },
	extra padding/.default=0.1cm,
}
\pgfplotsset{width=8cm, 
					every axis/.append style={                    
                    axis line style={<->}, 
                    label style={font=\tiny},
                    tick label style={font=\tiny}
                    }}

\begin{tikzpicture}[extra padding]
\begin{axis}[
  title={},
  xlabel={Solved instances},
  ylabel={Time},
  axis x line*=none,
  axis y line=box, 
  legend style={at={(0.35,+5.3cm)}, 
  anchor=north,legend columns=1, font = \tiny},
  nodes near coords,
  point meta=explicit symbolic,
]
\addplot+[mark size=1pt] coordinates {  (1, 20.14) [ ]
	(2, 21.80) [ ]
	(3, 24.36) [ ]
	(4, 30.37) [ ]
	(5, 35.91) [ ]
	(6, 51.12) [ ]
	(7, 55.86) [ ]
	(8, 55.89) [ ]
	(9, 58.90) [ ]
	(10, 63.23) [ ]
	(11, 71.85) [ ]
	(12, 128.37) [ ]
	(13, 237.20) [ ]
	(14, 271.17) [ ]
	(15, 392.09) [ ]
	(16, 398.46) [ ]
	(17, 829.54) [ ]
	(18, 908.66) [ ]
	(19, 1042.48) [ ]
	(20, 1050.24) [ ]
	(21, 1813.96) []
};
\addplot+[mark size=1pt] coordinates {  (1, 9.40) [ ]
	(2, 17.74) [ ]
	(3, 18.11) [ ]
	(4, 18.25) [ ]
	(5, 24.04) [ ]
	(6, 24.06) [ ]
	(7, 40.85) [ ]
	(8, 47.73) [ ]
	(9, 80.45) [ ]
	(10, 92.26) [ ]
	(11, 113.28) [ ]
	(12, 115.68) [ ]
	(13, 122.02) [ ]
	(14, 129.86) [ ]
	(15, 228.11) [ ]
	(16, 431.24) [ ]
	(17, 615.94) [ ]
	(18, 1283.69) [ ]
	(19, 1795.59) [ ]
	(20, 1798.68) [ ]
	(21, 1803.24) []
};

\addplot+[mark size=1pt] coordinates {  (1, 7.26) [ ]
 (2, 7.58) [ ]
 (3, 8.23) [ ]
 (4, 8.97) [ ]
 (5, 10.11) [ ]
 (6, 11.60) [ ]
 (7, 14.57) [ ]
 (8, 36.17) [ ]
 (9, 45.59) [ ]
 (10, 71.66) [ ]
 (11, 97.45) [ ]
 (12, 110.54) [ ]
 (13, 176.52) [ ]
 (14, 263.96) [ ]
 (15, 453.58) [ ]
 (16, 487.80) [ ]
 (17, 851.20) [ ]
 (18, 1820.57) [ ]
 (19, 1822.30) [ ]
 (20, 1823.19) [ ]
 (21, 1827.16) []
 };
\addplot+[mark size=1pt] coordinates {  (1, 15.11) [ ]
 (2, 15.92) [ ]
 (3, 27.75) [ ]
 (4, 36.78) [ ]
 (5, 40.69) [ ]
 (6, 48.22) [ ]
 (7, 62.40) [ ]
 (8, 68.67) [ ]
 (9, 101.32) [ ]
 (10, 154.82) [ ]
 (11, 203.07) [ ]
 (12, 434.85) [ ]
 (13, 673.28) [ ]
 (14, 835.17) [ ]
 (15, 888.45) [ ]
 (16, 936.20) [ ]
 (17, 1000.09) [ ]
 (18, 1195.11) [ ]
 (19, 1799.80) [ ]
 (20, 1811.15) [ ]
 (21, 1812.37) []
 };

\addplot+[mark size=1pt] coordinates {  (1, 8.55) [ ]
 (2, 13.14) [ ]
 (3, 14.46) [ ]
 (4, 18.13) [ ]
 (5, 18.88) [ ]
 (6, 22.77) [ ]
 (7, 26.16) [ ]
 (8, 39.20) [ ]
 (9, 93.11) [ ]
 (10, 272.08) [ ]
 (11, 911.86) [ ]
 (12, 952.98) [ ]
 (13, 1087.09) [ ]
 (14, 1287.25) [ ]
 (15, 1520.72) [ ]
 (16, 1605.87) [ ]
 (17, 1666.86) [ ]
 (18, 1741.99) [ ]
 (19, 1798.38) [ ]
 (20, 1801.98) [ ]
 (21, 1806.07) []
 };

	\legend{ -pythagorean, -pythagorean -dyn-sym-exploit, -static-sym-breaking, -dyn-sym-exploit, -native};
\end{axis}
\end{tikzpicture}
\caption{Results for the Pythagorean-triples benchmarks.}
	\label{fig:pyth}
\end{figure}
\begin{table*}
\begin{center}
		\begin{tabular}{lcccccccc} \toprule
\textbf{Configuration}& \textbf{Timed-}	& \textbf{Time}	&\textbf{Time}	& \textbf{Conflicts} & \textbf{\eclauses} & \textbf{Over-} & \textbf{Active}& \textbf{Active}\\ 
&\textbf{out}&&\textbf{(par-2)}&&&\textbf{head}& -E-& -C- \\\midrule
-pythagorean 			& 1 	& 360.1 & 446.5	& 1,973,404 	& 60303.3 	& 0.2 	& 0.006 & 0.006 \\ 
-pythagorean -dyn-sym-exploit  		& 3 	& 419.5 & 676.6 & 1,864,767 	& 55264.4 	& 36.0 	& 0.007 & 0.006\\  
-static-sym-breaking 	& 4 	& 474.1 & 821.4	& 3,132,118 	& 0 	  	& 0		& 	  	& 		\\ 
-dyn-sym-exploit 		& 3 	& 579.1 & 837.4 & 2,558,436 	& 388.4 	& 50.5 	& 0.004	& 0.007 \\ 
(native) 				& 3 	& 795.6 & 1053.7& 3,901,308 	& 0			& 0		& 		& 0.007 \\  \midrule
-pythagorean			& 4		& 578.0 & 1008.9& 3,045,218.4 	& 214,993.7	&	0.4 & 0.004	& 0.054 \\ 
-dyn-sym-exploit 		& 9		& 931.7	& 1714.1& 1,813,843.8 	& 3,541,747.0& 568.9 & 0.005 &	0.007 \\ \bottomrule
	\end{tabular}
\caption{Results for the Boolean Pythagorean triples problem, over 21 benchmarks. The bottom two configurations are without filtering.
} \label{tab:pyth}
\end{center}
\end{table*}

\section{Conclusions and future work}
We presented a general condition for adding what we call \eclauses, right after conflict analysis. We showed how this technique generalizes `symmetry' and `almost symmetry', and that indeed this method can add far more clauses than dynamic symmetry exploitation and related methods that are solely based on such symmetries. We showed several known problems for which this is relevant, and mentioned cases in which it was already done in the past with empirical success. 

There are three lines of future work that we consider important. 
First, it is important to classify additional problems as having the property that they are amenable to adding \eclauses, and check whether it can assist in accelerating their solving. Second, we foresee a dedicated SAT solver that maintains and reasons about \emph{clause generators}. That is, instead of adding many \eclauses as normal clauses, just keep the base learned clause with its bounds. It can be faster than the alternative of adding all \eclauses and does not suffer from the necessity to delete most of them. 
In a sense, this way the \eclauses are generated lazily, on demand, and then immediately erased. 
There are many implementation details that need to be developed for this. For example, one can add the generator to the watch list of all the literals that would have watched one of its generated \eclauses. In BCP, that literal tells us how to apply the unit implication rule to the generator. The reason clause can be maintained as a pair of a reference to the generator and an instantiation index. Many other details still need to be worked out. 

A third direction, is to control the BCP order, such that it works first on `normal' clauses and only if it terminates without a conflict, continue to propagate through the \eclauses, based on the assumption that the latter are less likely to cause a conflict at the current branch. One can also envision a SAT solver that splits BCP on normal and \eclauses between two threads. A possible high-level architecture is one in which the main thread, $T$, works on `normal' clauses and then on \eclauses, and the other, $T_e$, in the other direction. The first that finds a conflict terminates the other, or, alternatively, the solver chooses the better conflict clause based on its LBD and backtracking level.

\bibliographystyle{plain}
\bibliography{biblio}
\end{document}